\documentclass[apj]{emulateapj}

\newcommand{\kms}{\ifmmode{~{\rm km~s^{-1}}}\else{~km s$^{-1}$}\fi}
\newcommand{\cubecm}{\ifmmode{~{\rm cm^{-3}}}\else{~cm$^{-3}$}\fi}
\newcommand{\lsim}{\lower0.3em\hbox{$\,\buildrel <\over\sim\,$}}
\newcommand{\gsim}{\lower0.3em\hbox{$\,\buildrel >\over\sim\,$}}

\newcommand{\enzo}{{\sl Enzo}}
\newcommand{\Ms}{\ifmmode{M_\odot}\else{$M_\odot$}\fi}
\newcommand{\hh}{H$_2$}
\newcommand{\Ol}{$\Omega_\Lambda$}
\newcommand{\Om}{$\Omega_M$}
\newcommand{\Ob}{$\Omega_b$}
\newcommand{\mturb}{$\mathcal{M}_{turb}$}
\newcommand{\theat}{$t_{\rm{heat}}$}
\newcommand{\tcool}{$t_{\rm{cool}}$}
\newcommand{\rcool}{$r_{\rm{cool}}$}

\newcommand{\tdyn}{$t_{\rm{dyn}}$}

\newcommand{\tvir}{$T_{\rm{vir}}$}
\newcommand{\rvir}{$r_{\rm{vir}}$}
\newcommand{\rr}{$r_{200}$}
\newcommand{\lya}{Ly$\alpha$}
\newcommand{\zhh}{z$_{\rm{c,H2}}$}
\newcommand{\zlya}{z$_{\rm{c,Ly\alpha}}$}
\newcommand{\hv}{ {\bf h}}
\newcommand{\uv}{ {\bf u}}
\newcommand{\rv}{ {\bf r}}
\newcommand{\vv}{ {\bf v}}
\newcommand{\xv}{ {\bf x}}
\newcommand{\Sv}{ {\bf S}}

\begin{document}

\shorttitle{RESOLVING THE FORMATION OF PROTOGALAXIES I}
\shortauthors{WISE \& ABEL}

\title{Resolving the Formation of Protogalaxies. I. Virialization}
\author{John H. Wise and Tom Abel}
\affil{Kavli Institute for Particle Astrophysics and Cosmology,
  Stanford University, 2575 Sand Hill Road, MS 29, Menlo
  Park, CA 94025}
\email{jwise, tabel@slac.stanford.edu}

\begin{abstract}

  Galaxies form in hierarchically assembling dark matter halos. With
  cosmological three dimensional adaptive mesh refinement simulations,
  we explore in detail the virialization of baryons in the concordance
  cosmology, including optically thin primordial gas cooling. We focus
  on early protogalaxies with virial temperatures of 10$^4$ K and
  their progenitors.  Without cooling, virial heating occurs in shocks
  close to the virial radius for material falling in from
  voids. Material in dense filaments penetrates deeper to about half
  that radius. With cooling the virial shock position shrinks and also
  the filaments reach scales as small as a third the virial
  radius. The temperatures in protogalaxies found in adiabatic
  simulations decrease by a factor of two from the center and show
  flat entropy cores. In cooling halos the gas reaches virial
  equilibrium with the dark matter potential through its turbulent
  velocities. We observe turbulent Mach numbers ranging from one to
  three in the cooling cases. This turbulence is driven by the large
  scale merging and interestingly remains supersonic in the centers of
  these early galaxies even in the absence of any feedback
  processes. The virial theorem is shown to approximately hold over 3
  orders of magnitude in length scale with the turbulent pressure
  prevailing over the thermal energy. The turbulent velocity
  distributions are Maxwellian and by far dominate the small rotation
  velocities associated with the total angular momentum of the
  galaxies. Decomposing the velocity field using the Cauchy-Stokes
  theorem, we show that ample amounts of vorticity are present around
  shocks even at the very centers of these objects.  In the cold flow
  regime of galaxy formation for halo masses below 10$^{12} \Ms$, this
  dominant role of virialization driven turbulence should play an
  important role in for star formation as well as the build up of
  early magnetic fields.

\end{abstract}

\submitted{Submitted to ApJ on March 8, 2007}

\section{INTRODUCTION}

The process of virialization is clearly fundamental to all scales of
galaxy formation.  \citet{LB67} demonstrated that violent relaxation
occurs during the virialization of a dissipationless system, but does
the equivalent occur for the baryonic matter?  If it does, how it
achieves virial equilibrium should be inherently different because of
hydrodynamical effects and radiative cooling.  Additionally, this
would create a Maxwellian velocity distribution for the baryons as
well.  Turbulent velocities would exceed rotational ones.  This would
be at odds with the standard galaxy formation theories, which
generally assume smooth solid body rotating gaseous distributions
\citep[e.g.][]{Crampin64, Fall80, Mo98}.  The first occurrence of
widespread star formation can be regarded as the commencement of
galaxy formation, and its feedback on its host will affect all
subsequent star formation.  It is crucial to model the initial stage
of galaxy formation accurately.  Differences in initial configurations
of a collapsing halo may manifest itself in different types of central
luminous objects, whether it be a stellar disk \citep{Fall80, Mo98}, a
starburst \citep[see \S4 in][for a review]{Kennicutt98}, or a massive
black hole \citep{Bromm03, Volonteri05, Spaans06, Begelman06}.  These
differences may result from varying merger histories and the ensuing
virial heating or turbulence generation of the new cosmological halo.

For galaxy clusters, cosmological virialization has been studied
extensively \citep{Norman99, Nagai03a, Nagai03b, Schuecker04,
  Dolag05}.  It is customary to connect the velocity dispersion to a
temperature through the virial theorem for a collisionless system,
where the potential energy equals twice the kinetic energy.  However
X-ray observations and such cosmological simulations of galaxy
clusters have indicated that turbulent energies are comparable to
thermal energies.  Central turbulent pressure decreases the density,
but the temperature is largely unchanged.  This leads to an increased
entropy and a flatter entropy radial profile \citep{Dolag05} that is
in better agreement with X-ray observations \citep[e.g.][]{Ponman99}.
Simulations of merger dynamics suggest that turbulence is mostly
generated in Kelvin-Helmholtz instabilities between bulk flows and
virialized gas during minor mergers \citep[e.g.][]{Ricker01,
  Takizawa05}.  Alternatively turbulence can be generated by
conduction \citep{Kim03, Dolag04} or acoustic transport of energy
\citep{Norman99, Cen05}.

In standard galaxy formation models \citep{Rees77, Silk77, White78,
  Blumenthal84, White91, Mo98}, gas shock-heats to the virial
temperature as it falls into DM halos.  These models succeed with
considerable accuracy in matching various observables, such as star
formation histories, galaxy luminosity functions, and the Tully-Fisher
relationship \citep{White91, Lacey91, Cole94, Cole00}.  Galaxy
formation models depend on the virial temperature, most notably
through the cooling function that controls star formation rates and
their associated feedback mechanisms.  Atomic hydrogen and helium
radiative cooling is efficient in halos with masses between $10^8$ and
$10^{12} \Ms$ as cooling times can be less than the dynamical time of
the system, a condition that galaxy clusters do not satisfy.  This
strong cooling suggests that in galaxies thermal energy may be less
important for virialization than turbulent kinetic energy.  Motivated
by the results of galaxy cluster turbulence, we investigate this
potentially important role of turbulence and radiative cooling in
galaxy formation, using a series of high resolution numerical
simulations of protogalactic halos in this work.

We consider kinetic energy and pressure forces in our virial analysis
of protogalactic halos \citep[cf.][]{Shapiro99, Iliev01}.  This allows
us to investigate the equilibrium throughout the entire halo and
determine the importance of each energy component in the virial
theorem.  Kinetic energy can be decomposed into radial and azimuthal
motions along with turbulence, which affects the collapse of gas
clouds primarily in three ways. First as seen in galactic molecular
clouds, turbulence plays an integral part in current theories of star
formation as the density enhancements provide a favorable environment
for star formation \citep{Larson79, Larson81, Myers99, Goldman00}.
Second if the turbulence is supersonic, gas dissipates kinetic
energy through radiative cooling, which aids the gaseous collapse
\citep{Rees77}.  Conversely turbulent pressure adds an additional
force for the collapsing object to overcome and can delay the collapse
into a luminous object.  Last, turbulence provides an excellent
channel for angular momentum transport as the halo settles into
rotational equilibrium to satisfy Rayleigh's inviscid rotational
stability argument \citep{Rayleigh20, Chandra61} in which the specific
angular momentum must increase with radius.  Cosmological hydrodynamic
simulations have just begun to investigate angular momentum transport
within turbulent collapsing objects, and turbulence seems to play a
large role in segregating low (high) angular momentum gas to small
(large) radii \citep{Norman99, Abel02, Yoshida06b}.

We study idealized cases of structure formation where stellar feedback
is ignored because it provides a convenient problem to focus on the
interplay between cosmological merging, hydrodynamics, and cooling
physics during the assembly of early halos.  Some of the discussed
physical principles should, however, be applicable to galaxies of all
masses.  These simulations provide the simplest scenario to which we
can incrementally consider further additional physics, such as \hh~and
HD cooling physics \citep{Saslaw67, Palla83, Flower00}, primordial
stellar feedback \citep{Whalen04, Kitayama04, Alvarez06, Yoshida06a,
  Abel07}, metal enrichment from primordial stars \citep{Heger02,
  Tumlinson06}, AGN feedback \citep{Springel05, Kuhlen05}, and
``normal'' metal-enriched star formation \citep[see][for a
review]{Larson03}.

We present a suite of adaptive mesh refinement simulations that are
described in \S2.  Then we analyze the local virial equilibrium and
shocks in halos in \S3.  There we also differentiate between infall
through voids and filaments and its associated virialization.  We
discuss the situations in which virial heating and turbulence occur.
Next in \S4, we decompose the velocity distribution in principle axes
to explore virialization in both the DM and baryonic components.
Furthermore we decompose velocities into shear and compressible flows
to study turbulent flows in the virialized gas.  We discuss the
implications of these results on star and galaxy formation in \S5.
Finally we summarize in the last section.

%
%

\begin{deluxetable*}{ccccccc}
\tablecolumns{7}
\tabletypesize{}
\tablewidth{0pc}
\tablecaption{Simulation Parameters\label{tab:sims}}

\tablehead{
  \colhead{Name} & \colhead{l} & \colhead{z$_{end}$} & \colhead{N$_{part}$} &
  \colhead{N$_{grid}$} & \colhead{N$_{cell}$} & \colhead{Cooling model} \\
  \colhead{} & \colhead{[Mpc]} & \colhead{} & \colhead{} & \colhead{}
  & \colhead{} & \colhead{}
}
\startdata

A0 & 1.0 & 15.87 & 2.22 $\times$ 10$^7$ & 30230 & 9.31 $\times$ 10$^7$
(453$^3$) & Adiabatic \\

A6 & 1.0 & 15.87 & 2.22 $\times$ 10$^7$ & 40486 & 1.20 $\times$ 10$^8$
(494$^3$) & H,He \\

A9 & 1.0 & 18.74 & 2.22 $\times$ 10$^7$ & 45919 & 1.21 $\times$ 10$^8$
(495$^3$) & H,He,\hh \\

B0 & 1.5 & 16.80 & 1.26 $\times$ 10$^7$ & 23227 & 6.47 $\times$ 10$^7$
(402$^3$) & Adiabatic \\

B6 & 1.5 & 16.80 & 1.26 $\times$ 10$^7$ & 21409 & 6.51 $\times$ 10$^7$
(402$^3$) & H,He \\

B9 & 1.5 & 23.07 & 1.26 $\times$ 10$^7$ & 20525 & 5.59 $\times$ 10$^7$
(382$^3$) & H,He,\hh

\enddata
\tablecomments{Col. (1): Simulation name. Col. (2): Comoving box
  size. Col. (3): Final redshift. Col. (4): Number of dark matter
  particles. Col. (5): Number of AMR grids at the final
  redshift. Col. (6): Maximum number of unique AMR grid
  cells. Col. (7): Cooling model.}
\end{deluxetable*}

\section{THE SIMULATIONS}

To investigate protogalactic (\tvir~$>10^4$ K) halo virialization in
the early universe, we utilize an Eulerian structured, adaptive mesh
refinement (AMR), cosmological hydrodynamical code, \enzo\footnote{See
  http://http://lca.ucsd.edu/portal/software/enzo} \citep{Bryan97,
  Bryan99, OShea04}.  \enzo~solves the hydrodynamical equations using
the second order accurate piecewise parabolic method
\citep{Woodward84, Bryan95}, while a Riemann solver ensures accurate
shock capturing with minimal viscosity.  Additionally \enzo~uses an
adaptive particle-mesh $n$-body method to calculate the dynamics of
the collisionless dark matter particles \citep{Couchman91}.  Regions
of the simulation grid are refined by two when one or more of the
following conditions are met: (1) Baryon density is greater than 3
times $\Omega_b \rho_0 N^{l(1+\phi)}$, (2) DM density is greater than
3 times $\Omega_{\rm{CDM}} \rho_0 N^{l(1+\phi)}$, and (3) the local
Jeans length is less than 4 cell widths.  Here $N = 2$ is the
refinement factor; $l$ is the AMR refinement level; $\phi = -0.3$
causes more frequent refinement with increasing AMR levels,
i.e. super-Lagrangian behavior; $\rho_0 = 3H_0^2/8\pi G$ is the
critical density; and the Jeans length, $L_J = \sqrt{15kT/4\pi\rho G
  \mu m_H}$, where $H_0$, $k$, T, $\rho$, $\mu$, and $m_H$ are the
Hubble constant, Boltzmann constant, temperature, gas density, mean
molecular weight in units of the proton mass, and hydrogen mass,
respectively.  The Jeans length refinement ensures that we meet the
Truelove criterion, which requires the Jeans length to be resolved by
at least 4 cells on each axis \citep{Truelove97}.

We conduct the simulations within the concordance $\Lambda$CDM model
with WMAP first year parameters (WMAP1) of $h$ = 0.72, \Ol~= 0.73,
\Om~= 0.27, \Ob~= 0.024$h^{-2}$, and a primordial scale invariant ($n$
= 1) power spectrum with $\sigma_8$ = 0.9 \citep{Spergel03}.  $h$ is
the Hubble parameter in units of 100 km s$^{-1}$ Mpc$^{-1}$.  \Ol,
\Om, and \Ob~are the fractions of critical energy density of vacuum
energy, total matter, and baryons, respectively.  Last $\sigma_8$ is
the rms of the density fluctuations inside a sphere of radius
8$h^{-1}$ Mpc.

Using the WMAP1 parameters versus the significantly different WMAP
third year parameters \citep[WMAP3;][]{Spergel06} have no effect on
the evolution of individual halos as are considered here.  However
these changes play an important role in statistical properties.  For
example, halos with mass $10^6 \Ms$ at redshift 20 correspond to
$2.8\sigma$ peaks with the WMAP1 but are $3.5\sigma$ peaks for WMAP3.
The \Om/\Ob~ratio also only changed from 6.03 to 5.70 in WMAP3.  

We also have verified that there is nothing atypical about the mass
accretion rate histories of the objects we study.  The mass accretion
history of these objects exhibit smooth growth during minor mergers
and accretion and dramatic increases when a major merger occurs.  This
behavior is consistent with typical halo assemblies in extended
Press-Schechter calculations \citep{Bond91, Bower91, Lacey93, Lacey94,
  vdBosch02b} and cosmological numerical simulations
\citep[e.g.][]{DeLucia04, Gao05}.  The mass accretion histories in our
simulations are well described by the fitting function of
\citeauthor{vdBosch02b} with $M_0 = 3 \times 10^7 \Ms$, $z_f = 17$,
and $\nu = 12.5$.  We also compare our data against the mass accretion
histories of \citeauthor{Gao05}, who tested their data against an
extended Press-Schechter calculation of the growth history of the
halos.  We find no major discrepancies between the two histories.

The initial conditions of this simulation are well-established by the
primordial temperature fluctuations in the cosmic microwave background
(CMB) and big bang nucleosynthesis (BBN) \citep[][and references
therein]{Hu02, Burles01}.

We perform two realizations with different box sizes and random
phases.  In the first simulation (simulation A), we set up a
cosmological box with 1 comoving Mpc on a side, periodic boundary
conditions, and a 128$^3$ top grid with three nested child grids of
twice finer resolution each.  The other simulation is similar but with
a box side of 1.5 comoving Mpc (simulation B).  We provide a summary
of the simulation parameters in Table \ref{tab:sims}.  These volumes
are adequate to study halos of interest because the comoving number
density of $>$10$^4$ K halos at $z=10$ is $\sim$6 Mpc$^{-3}$ according
to an ellipsoidal variant of Press-Schechter formalism
\citep{Sheth02}.  We use the COSMICS package to calculate the initial
conditions at $z$ = 129 (119)%
\renewcommand{\thefootnote}{\fnsymbol{footnote}}%
\footnote{To simplify the discussion, simulation A will always be
  quoted first with the value from simulation B in parentheses.}%
\renewcommand{\thefootnote}{\arabic{footnote}} %
\citep{Bertschinger95, Bertschinger01}, which calculates the linear
evolution of matter fluctuations.  We first run a dark matter
simulation to $z=10$ and locate the DM halos using the HOP algorithm
\citep{Eisenstein98}.  We identify the first dark matter halo in the
simulation with \tvir~$>$ 10$^4$ K and generate three levels of
refined, nested initial conditions with a refinement factor of two,
centered around the Lagrangian volume of the halo of interest.  The
nested grids that contain finer grids have 8 cells between its
boundary and its child grid.  The finest grid has an equivalent
resolution of a 1024$^3$ unigrid.  This resolution results in a DM
particle mass of 30 (101) $\Ms$ and an initial gas resolution of 6.2
(21) $\Ms$.

\enzo~employs a non-equilibrium chemistry model \citep{Abel97,
  Anninos97}.  We conduct three simulations for each realization with
(i) the adiabatic equation of state with an adiabatic index $\gamma =
5/3$, (ii) a six species chemistry model (H, H$^{\rm +}$, He, He$^{\rm
  +}$, He$^{\rm ++}$, e$^{\rm -}$), and (iii) a nine species chemistry
model that adds H$_2$, H$_2^{\rm +}$, and H$^{\rm -}$ to the six
species model.  In the nine species model, we use the molecular
hydrogen cooling rates from \citet{Galli98}.  These models are
differentiated in the text by denoting 0, 6, and 9, respectively,
after the simulation name (e.g. simulation B0).  Compton cooling and
heating of free electrons by the CMB and radiative losses from atomic
and molecular cooling are also computed in the optically thin limit.

To restrict the analysis to protogalactic halos in the \hh~models, we
suppress \hh~formation in halos that cannot undergo \lya~cooling by
reducing the residual electron fraction to $10^{-12}$ instead of a
typical value of $\sim10^{-4}$ only at the initial redshift
\citep{Shapiro94}.  This mimics an extreme case where all \hh~is
dissociated by an extremely large radiation background, and the halo
can only collapse and form stars when free electrons from ionized
hydrogen can catalyze \hh~formation.

We end the simulations with non-equilibrium cooling when the gas
begins to rapidly cool and collapse.  We choose a final resolution
limit of $\sim$3000 (4000) proper AU, corresponding to a refinement
level of 15.  We end the adiabatic simulations at the same redshift.
In a later paper, we will address the collapse of these halos to much
smaller scales.

\section{VIRIAL ANALYSIS}

The equation of motion for an inviscid gas in tensor notation reads:
\begin{equation}
\rho \frac{D v_i}{Dt} = -\frac{\partial}{\partial x_i} p  + \rho g_i \nonumber
\label{mest3.39}
\end{equation}
where $D/Dt = \partial /\partial t + v_j \partial / \partial x_j $ is
the total derivative. Here $v$ is velocity; $p$ is pressure; $\rho$ is
density; and $g = \nabla\Phi$ where $\Phi$ is the gravitational
potential.  From this \citet{Chandra53} derived the general virial
theorem for a region contained within a surface $\Sv$, in scalar form,
\begin{equation}
  \frac{1}{2} \frac{D^2 I }{Dt^2} =  2 \mathcal{T} + \mathcal{V}
  + 3 (\gamma-1)\mathcal{E} - \int p  \, \rv \cdot   d\Sv,
\label{vt}
\end{equation}
where 
\begin{equation}
  \mathcal{V} = -\frac{1}{2}G \int_V
  \frac{\rho(\xv)\rho(\xv')}{|\xv - \xv' |} d\xv d\xv' ,
\end{equation}
$\mathcal{T} = \frac{1}{2} \int \rho \vv^2 d\xv $, $I = \int \rho
\xv^2 d\xv$, and $\mathcal{E}=\int \varepsilon \; d\xv $ denote the
gravitational potential energy, the trace of the kinetic energy and
inertia tensor, and the total internal thermal energy, respectively.
The surface term $E_s$ (the last term in eq. [\ref{vt}]) is often
negligible in the outer regions of the halo.  The system is not
necessarily in virial equilibrium if $\ddot{I} = 0$, but the
time-averaged quantity is zero when the entire system is in virial
equilibrium.  A system is expanding or contracting whether $\ddot{I}$
is positive or negative, respectively, based on energy arguments.
\citet{Ballesteros06} gives counterexamples to this simple
interpretation.  However, in the cases presented here, spherically
averaged radial velocities are always negative.

%
%

\begin{deluxetable}{ccccccc}
\tablecolumns{7}
\tabletypesize{}
\tablewidth{0pc}
\tablecaption{Halo Properties\label{tab:halos}}

\tablehead{
  \colhead{Name} & \colhead{z$_{coll}$} & \colhead{M$_{tot}$}
  & \colhead{$\rho_c$} &
  \colhead{\tvir\tablenotemark{a}} & \colhead{T$_c$} &
  \colhead{$\langle T \rangle$} \\
  \colhead{} & \colhead{} & \colhead{[$\Ms$]} &
  \colhead{[cm$^{-3}$]} & \colhead{[K]} &
  \colhead{[K]} & \colhead{[K]}
}
\startdata
\multicolumn{7}{c}{\hh~Induced Collapse}\\
\noalign{\vskip 3pt \hrule \vskip 3pt}
A0 & 18.74    & 9.8 $\times$ 10$^6$ & 8.1 & 9200 & 10000 & 5700 \\
A6 & 18.74    & 9.8 $\times$ 10$^6$ & 17 & 9200 & 7700 & 5500 \\
A9 & 18.74    & 9.8 $\times$ 10$^6$ & 1.6 $\times$ 10$^6$ & 9200 & 590 & 5000 \\

B0 & 23.07    & 6.2 $\times$ 10$^6$ & 13 & 8300 & 12000 & 6000 \\
B6 & 23.07    & 6.2 $\times$ 10$^6$ & 15 & 8300 & 8900 & 5600 \\
B9 & 23.07    & 6.7 $\times$ 10$^6$ & 3.0 $\times$ 10$^6$ & 8700 & 580 & 4200 \\

\noalign{\vskip 3pt \hrule \vskip 3pt}
\multicolumn{7}{c}{\lya~Induced Collapse} \\
\noalign{\vskip 3pt \hrule \vskip 3pt}
A0 & 15.87 & 3.6 $\times$ 10$^7$ & 4.9 & 19000 & 17000 & 12000 \\
A6 & 15.87 & 3.6 $\times$ 10$^7$ & 1.8 $\times$ 10$^6$ & 19000 & 8700 & 7300 \\

B0 & 16.80 & 3.5 $\times$ 10$^7$ & 3.8 & 19000 & 31000 & 11000 \\
B6 & 16.80 & 3.6 $\times$ 10$^7$ & 4.0 $\times$ 10$^6$ & 20000 & 9000 & 7500

\enddata
\tablenotetext{a}{Virial temperatures are calculated with $\mu$ = 1.22
  in all cases.}
\tablecomments{Col. (1): Simulation name. Col. (2): Redshift of
  collapse through \hh~or \lya~cooling. Col. (3): Total
  mass. Col. (4): Central density. Col. (5): Virial temperature
  (i.e. eq. [\ref{eqn:tvir}]). Col. (6): Central
  temperature. Col. (7): Mass-averaged temperature of the entire
  halo.}
\end{deluxetable}

We define the halo as the material contained in a sphere with a radius
$r_{200}$ enclosing an average DM overdensity of 200 and as such
relates to mass by
\begin{equation}
\label{eqn:r200}
r_{200} = \left[ \frac{GM}{100\Omega_{\rm{CDM}}(z) H^2(z)} \right]^{1/3},
\end{equation}
where $M$ is the mass of the halo, $\Omega_{\rm{CDM}}(z)$ is evaluated
at a redshift $z$, and $H$ is the Hubble parameter at $z$.  The region
where the cooling time is shorter than a Hubble time is denoted as the
cooling radius \rcool,
\begin{equation}
\label{eqn:rcool}
t_{\rm{cool}} (r_{\rm{cool}}) \equiv H(z)^{-1}
\end{equation}
\citep{White91}.  Mass and radius define a circular velocity and
virial temperature, which are
\begin{equation}
\label{eqn:tvir}
V_c = \sqrt{\frac{GM}{r_{200}}}
\quad \textrm{and} \quad
T_{vir} = \frac{\mu m_p V_c^2}{2k},
\end{equation}
for a singular isothermal sphere \citep[see][with $\beta$ = 1 and
$\Delta_c$ = 200]{Bryan98}.  0.59 and 1.22 in units of the proton mass
are appropriate values for $\mu$ for the fully ionized and completely
neutral states of a primordial hydrogen and helium mixture of gas,
respectively.  We use $\mu$ = 1.22 throughout this paper.  We note
that \citet{Iliev01} considered non-singular, truncated isothermal
spheres, and the resulting virial temperature is $\sim$15\% lower than
the one calculated in equation (\ref{eqn:tvir}).

For $\gamma = 5/3$, \tvir~is the temperature at which an ideal
adiabatic gas reaches virial equilibrium with the specified potential.
Please note that for an isothermal gas where $\gamma$ is close to
unity virial equilibrium is established between the turbulent energies
and the gravitational potential as the $3 (\gamma-1) \mathcal{E}$ term
in equation~(\ref{vt}) goes to zero.

%
%
\begin{figure*}
\begin{center}
\plotone{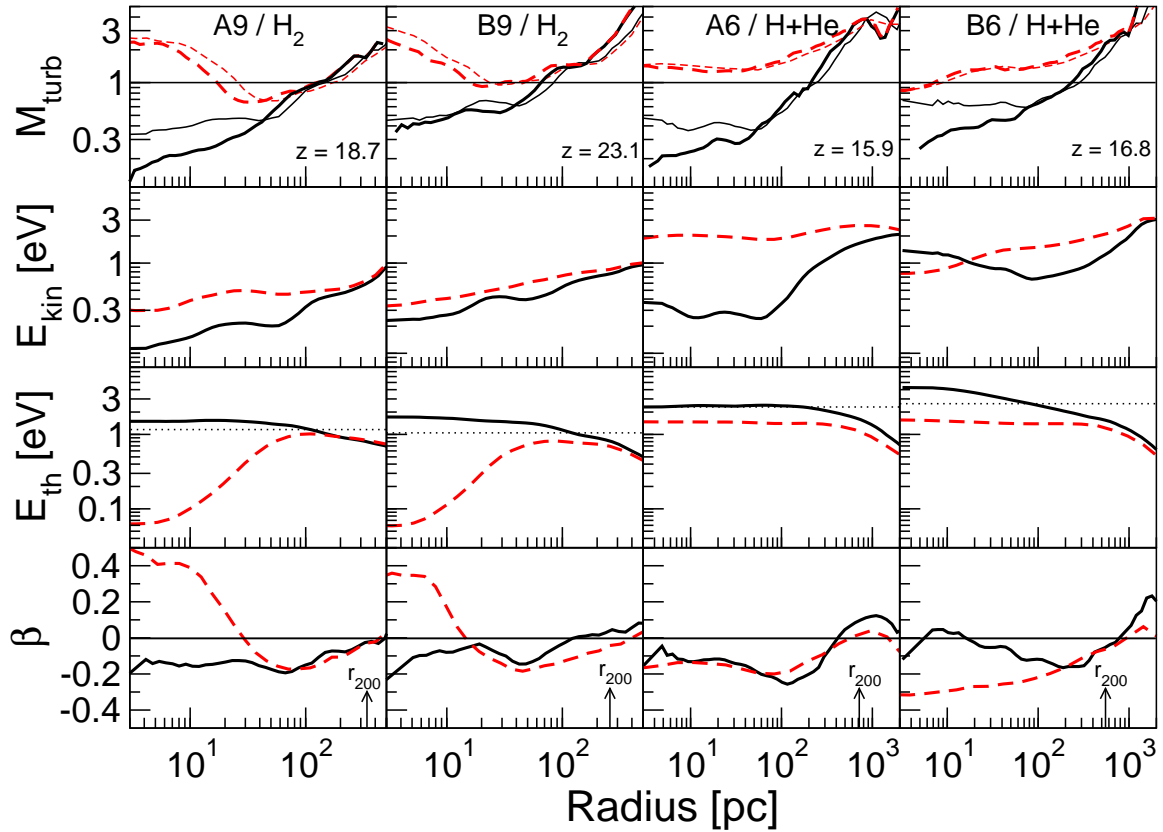}
\caption{\label{fig:energies} A comparison of (\textit{top to bottom})
  turbulent Mach numbers, turbulent and thermal energies, and virial
  parameters between simulations with radiative cooling
  (\textit{dashed}) and adiabatic models (\textit{solid}).  The main
  coolant is listed at the top of each column.  The \textit{first} and
  \textit{second} columns display the state of these variables at z =
  \zhh~= 18.7 (23.1) for simulation A and B, respectively.  The
  \textit{third} and \textit{fourth} columns are the data at z =
  \zlya~= 15.9 (16.8).  The \textit{top} row depicts the importance of
  radiative cooling in generating trans- and super-sonic turbulence
  throughout the halo during virialization.  The \textit{thick} lines
  represent the turbulent Mach number (eq. [\ref{eqn:mach}]), and the
  \textit{thin} lines show the Mach number using the rms velocity with
  respect to the mean velocity of the halo.  The \textit{middle} two
  rows show that when radiative cooling is efficient the halo cannot
  virialize through heating but must virialize by increasing its
  kinetic (turbulent) energies.  The dotted line in the \textit{third}
  row marks \tvir~(eq. [\ref{eqn:tvir}]) with $\mu$ = 1.22, estimated
  from the total halo mass.  We plot the virialization parameter
  $\beta$ (eq. [\ref{eqn:beta}]) to investigate the local virial
  equilibrium ($\beta = 0$), particularly at \rr.  Furthermore,
  $\beta$ allows us to determine the mass-averaged dynamics of the
  system at a given radius, where $\beta > 0$ and $< 0$ correspond to
  decelerating and accelerating collapses, respectively.
  \rr~(eq. [\ref{eqn:r200}]) is marked on the bottom of each column.}
\end{center}
\end{figure*}

%
%
\begin{figure}
\begin{center}
\plotone{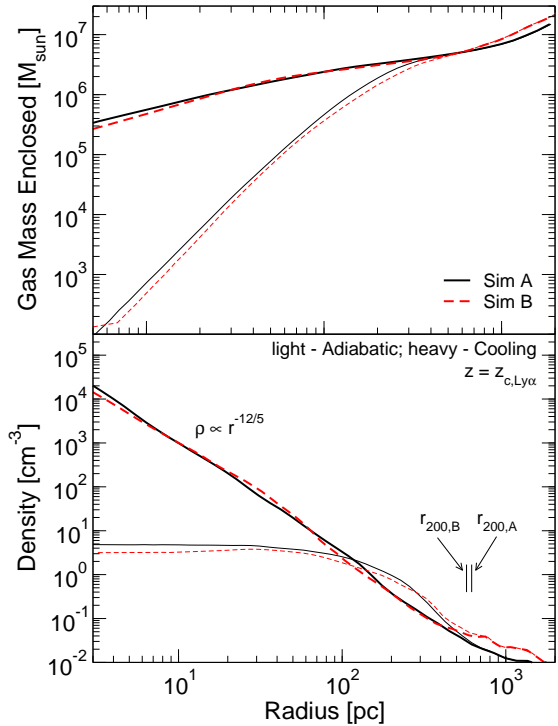}
\caption{\label{fig:dens} Mass-weighted radial profiles for gas mass
  enclosed (\textit{top}) and number density (\textit{bottom}) for
  simulations A (\textit{solid black}) and B (\textit{dashed red}) at
  z = \zlya~= 15.9 (16.8).  The \textit{light} and \textit{heavy}
  lines represent data for adiabatic and cooling models, respectively.
  The virial shock in the cooling halos occurs at $\frac{2}{3}$\rr,
  illustrated by the density increasing at smaller radii.  We mark the
  radius $r_{200}$ = 615 (576) for simulation A (B).}
\end{center}
\end{figure}

\subsection{Local Analysis}
\label{sec:local}

We evaluate the terms of equation (\ref{vt}) with respect to radius
(i.e. the volume contained in a radius $r$).  Figure
\ref{fig:energies} illustrates the radial structure of (a) the
turbulent Mach number,
\begin{equation}
  \label{eqn:mach}
  \mathcal{M}_{turb} = \frac{v_{rms}}{c_s}; \quad
  c_s = \sqrt{\frac{\gamma kT}{\mu m_h}}
\end{equation}
(b) turbulent and (c) thermal energies per $m_h$, and (d) a
``virialization'' parameter\footnote{This is a modified version of the
  $\beta$ used in \citet{Shaw06} to account for kinetic energies and
  so that it does not diverge when $\mathcal{V} \rightarrow 0$ in the
  center.  It still has the same behavior of $\beta \rightarrow 0$ as
  $\ddot{I} \rightarrow 0$.}
\begin{equation}
  \label{eqn:beta}
  \beta = \frac{3(\gamma - 1)\mathcal{E} + 2\mathcal{T}}{E_s -
    \mathcal{V}} - 1,
\end{equation}
of the adiabatic and radiative cooling simulations when the cooling
halo collapses.  Here $v_{rms}$ is the three-dimensional rms velocity
and is assessed using the gas velocities relative to the mean gas
velocity of each spherical shell.  In the top row of Figure
\ref{fig:energies}, we also plot the Mach number, using $v_{rms}$ with
respect to the mean velocity of the gas within $r_{200}$.  In the six
and nine species simulation, this occurs at \zhh~= 18.7 (23.4) and
\zlya~= 15.9 (16.8), respectively.  The radial profiles are centered
on the densest point in the simulation with the collapsing halo.
Several properties of the most massive halo in each simulation are
detailed in Table \ref{tab:halos}.  The sections of the Table compare
the halo in the adiabatic, \lya, and \hh~simulations.


Figure \ref{fig:dens} shows the mass-weighted radial profiles of gas
mass enclosed and gas density at z = \zlya~in the adiabatic and
cooling cases.  Both realizations are remarkably similar.  Halos in
the adiabatic case have a central core with a radius $\sim$50 pc and
gas density of $\sim$5.0 (3.5)\cubecm.  Core densities in simulation A
are slightly higher than simulation B, which has larger thermal and
turbulent pressures (see Figure \ref{fig:energies}).  With radiative
cooling, gas infalls rapidly as it cools and undergoes a self-similar
collapse with $\rho \propto r^{-12/5}$.

\subsubsection{Virial Radius}

We define the virial radius \rvir~when $\beta = 0$ and $d\beta/dr <
0$.  When we radially average the \lya~halo, \rvir~= 419 (787) pc in
the adiabatic simulations where the corresponding $r_{200}$ value is
615 (576) pc.

\begin{figure*}
\begin{center}
\plotone{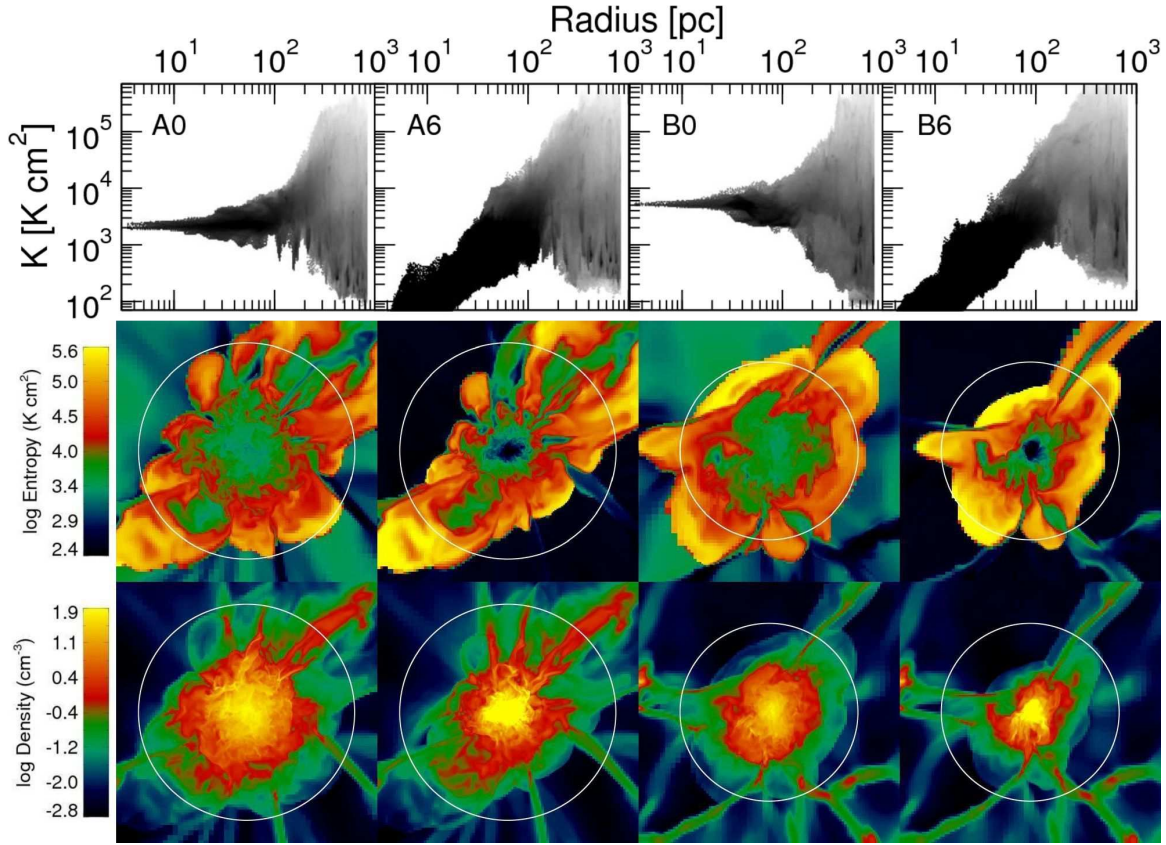}
\caption{\label{fig:entropy} Differences in entropy and density in a
  protogalactic halo at z = \zlya.  \textit{Left to right}:
  simulations A0, A6, B0, B6.  \textit{Top row}: Two-dimensional
  histograms of radius and the adiabatic invariant $K = T / n^{2/3}$.
  Darker pixels represent a higher mass for a particular radius and
  $K$.  This plot shows the wide variations in $K$ at $r > 100$ pc,
  where cold flows and shock-heated gas coexist at a given radius.
  Radiative cooling allows the gas to cool and collapse in the center
  that accounts for the decrease in simulations A6 and B6.  The
  material at $r \gsim 100$ pc and $K \lsim 10^{3.3}$ K cm$^2$
  corresponds to the cold flows inside filaments that illustrates that
  virialization occurs at different radii depending on its origin.
  \textit{Middle row}: Two-dimensional slices of entropy.  The circle
  denotes \rr = 615 pc and 576 pc for simulation A and B,
  respectively.  The virial shock exists at approximately \rr~in the
  adiabatic models; however it shrinks to $\sim$2/3 of \rr~when we
  consider radiative cooling.  \textit{Bottom row}: Two-dimensional
  slices of number density of baryons.}
\end{center}
\end{figure*}

A well defined shock exists between the interface between voids and
the halo.  This material shock-heats to \tvir~and virializes at a
radius comparable to \rr~in the adiabatic cases.  When we include
radiative cooling, this radius decreases everywhere around the
halo-void virial shock and is low as \rr/2. In contrast to
the voids, the filamentary gas shock-heats at an even smaller radius.
\citet{Dekel06} also studied the stability of cold inflows within a
hot virialized medium and found similar results.  Figures
\ref{fig:dens} and \ref{fig:entropy} illustrate these changes.  At
\rr, densities in the adiabatic case begin to increase more rapidly
than the cooling case as material accretes at the virial shock.  No
significant increase in $d\rho/dr$ is seen in the cooling case,
indicative of a self-similar collapse.

\subsubsection{Adiabatic Model}
\label{sec:adiabatic}

We start with the discussion of the adiabatic model as it is the
simplest case and later compare the calculations with radiative
cooling to this model.  Virialization should transfer potential energy
to kinetic energy that dissipates in shocks to thermal energy, which
is the implication of the dissipationless virial theorem.  The solid
lines in Figure \ref{fig:energies} represent the energies in adiabatic
models.  The physics illustrated in this Figure are as follows:

\medskip

1. \textit{Thermal energy}--- The gas shock-heats to the virial
temperature at the virial shock.  Virial heating continues with
decreasing radius as the surface term becomes significant in the
interior.  The resulting central temperature of the halo is 10000
(12000) K, which is 1.2 (1.5) \tvir, at a redshift of \zhh~when the
\hh~model collapses.  At the time (z = \zlya) of collapse caused by
\lya~cooling, the central temperature is 17000 (31000) K,
corresponding to 0.9 (1.6) \tvir.

\medskip

2. \textit{Kinetic energy}--- It increases along with the thermal
energy during virialization.  The gas is generally turbulent,
appearing as a velocity dispersion with a bulk radial inflow.  At \rr,
the kinetic energy is equivalent to the thermal energy,
$\mathcal{T}/\mathcal{E} \sim 1$.  This ratio steadily drops toward
the center, where $\mathcal{T} / \mathcal{E} \sim 1/3$.  This decrease
in kinetic energy is apparent in all the calculations except
simulation B at \zlya, increasing by a factor of two in the center.

\medskip

3. \textit{Turbulent Mach number}--- At \rr, the turbulent Mach number
\mturb~is maximal and varies from 1--3 in all adiabatic simulations.
\mturb~decreases to subsonic values $\sim$0.15 but never below in the
interior.  Note that \mturb~does not increase as the turbulent energy
towards the center in simulation B because of the also growing sound
speed there.

\medskip

4. \textit{Virialization parameter}--- Virial equilibrium is
quantified by the virialization parameter $\beta$, where the collapse
is retarding or accelerating when it is negative or positive,
respectively.  At \zhh~and \zlya~and in both simulations, $\beta$ is
within 20\% of being virialized ($\beta = 0$).  For comparison
purposes, this corresponds to a halo having 80\% of the required
velocity dispersion for virialization in the dissipationless case.  At
\rvir, $\beta$ is nearly zero which defines the virialized object.
Within \rvir, the values decrease to values around --0.1 but stays
$\lsim$ 0.

\medskip

Characteristics of turbulence in our adiabatic models are similar to
ones found in galaxy cluster simulations \citep{Norman99, Dolag05}.
Both groups find that turbulence provides $\sim$5--30\% of the total
pressure, i.e. $\mathcal{T} / (\mathcal{T} + \mathcal{E}$), in the
cluster cores.  Our protogalactic halos have $\sim$25\% of the
pressure in the turbulent form.  Also the galaxy clusters in
\citet{Norman99} have comparable Mach numbers of $\sim$1.6 at \rvir,
$\sim$0.5 at \rvir/3, and $\sim$0.3 in the core.  These similarities
suggest that virial turbulence is generated over a large range of mass
scales.

%
%

\subsubsection{\lya~Cooling Model}
\label{sec:lyacollapse}

At \zlya, halos in calculations with the H+He cooling model start to
rapidly collapse.  The dashed lines in the third and fourth columns of
Figure \ref{fig:energies} illustrate the energies of this model.

\medskip

1. \textit{Thermal energy}--- Compared to the adiabatic models, the
gas can radiatively cool through \lya~emission to T $\sim$ 8000 K
within \rcool~$\sim$ \rvir.  The entire halo is isothermal at this
equilibrium temperature.  Below this temperature, the cooling function
of pristine gas drops by several orders of magnitude, and the gas can
no longer cool efficiently.  The thermal energy is $\sim$65\% lower
than the adiabatic case.

\medskip

2. \textit{Kinetic energy}--- In response to the lesser thermal
energy, the system tends toward virial equilibrium by increasing
kinetic (turbulent) energy.  The gravitational potential and surface
terms do not appreciably change with the inclusion of radiative
cooling.  Turbulent energy within \rvir~increases as much as a factor
of 5 when compared to the adiabatic case.

\medskip

3. \textit{Turbulent Mach number}--- The changes in thermal and
kinetic energies equate to a increase of \mturb~by a factor of 2--3 to
values up to 1.5.  The turbulence is supersonic in all cases at the
virial shock, but when we include radiative cooling, this trait
emanates inward as the halo begins to rapidly cool.  When the central
core becomes gravitationally unstable, the entire halo is
supersonically turbulent.

\medskip

4. \textit{Virialization parameter}--- The increased kinetic
energies compensate for the loss in thermal energy and the halo
remains in a similar virial state.  This is apparent in the remarkably
similar radial characteristics of $\beta$ in the adiabatic and H+He
models of simulation A.

\subsubsection{\hh~Cooling Model}
\label{sec:h2collapse}

The collapses caused by \hh~cooling at z = \zhh~have very similar
dynamics as the halos described in the previous section.  The dashed
lines in the first and second columns of Figure \ref{fig:energies}
illustrate the energies of this model.

\medskip

1. \textit{Thermal energy}--- \hh~cooling is efficient down to 300 K,
so gas can depose a much larger fraction of its thermal energy.
Inside \rcool~$\sim$ 0.32 (0.19) \rvir, thermal energies are only 5\%
of the values in the adiabatic models.

\medskip

2. \textit{Kinetic energy}--- The turbulent energies must increase as
in the \lya~case, and they increase by 93\% (44\%) on average inside
\rcool.

\medskip

3. \textit{Turbulent Mach number}--- Similarly, \mturb~increases up to
a factor of 10 to become supersonic at values up to 3 throughout the
halo.  They are somewhat larger than the \lya~cases since \hh~can cool
to significantly lower temperatures than the virial temperature.

\medskip

4. \textit{Virialization parameter}--- The virial equilibrium of the
halos are also similar to the other models.  $\beta$ smoothly
transitions from nearly equilibrium at \rvir~to an increased radial
infall with $\beta$ = --0.2 at 70 pc.  Then it increases to 0.4 inside
10 pc, which corresponds to the gas decelerating from the rapid infall
as it encounters the central molecular cloud.

\subsubsection{Model Summary}

Baryons are close to virial equilibrium over three orders of magnitude
in length scale by gaining both thermal and kinetic energies
independent of cooling physics.  Central temperatures of the adiabatic
simulations are up to twice the nominal virial temperature.  Similar
to galaxy cluster studies, turbulence in the adiabatic model
contributes $\sim$25\% to the energy budget with Mach numbers $\sim$
0.3 in the center.  In cooling cases, atomic and molecular cooling
inhibit virialization through heating, therefore the object must
virialize by gaining kinetic energy up to five times the energy seen
in the adiabatic models.  This translates into the flow becoming
supersonically turbulent with Mach numbers ranging from one to three.

\subsection{Variations in the virial shock}
\label{sec:shock}

Using the adiabatic invariant, $K = T/n^{2/3}$, which we label
``entropy'', allows us to differentiate between gas accreting from
voids and filaments.  As a precaution, we note that $K$ is not an
invariant when $\gamma$ varies; however, this is not the case in our
simulations in which we permit molecular hydrogen cooling.  Here
molecular fractions remain low, $< 10^{-3}$, and $\gamma \approx 5/3$
even in the densest regions.  The top row of Figure~\ref{fig:entropy}
depicts the variance of $K$ with respect to radius in two-dimensional
histograms, where the intensity of each pixel represents the mass
having the corresponding $K$ and $r$.  The middle and bottom rows
display two-dimensional slices of $K$ and density, respectively,
through the densest point in the halo.  The virialized gas from the
voids has low density and does not significantly contribute to the
mass averaged radial profiles.  Figure \ref{fig:entropy} illustrates
this gas at $r \sim r_{200}$ and $K \gsim 10^{4.5}$ K cm$^{2}$.  The
gas in filaments has lower entropy than the rest of the halo at $r >$
150 pc and $K \lsim 10^{3.3}$ K cm$^{2}$.  In equation (\ref{vt}), the
pressure in the surface term is the constant at a given radius.  The
accreting, denser, unshocked gas in filaments has lower temperatures
than the more diffuse accreting gas.  The gas remains cool until it
shocks and mixes well inside \rvir~and as small as $\sim$\rvir/4 in
the most massive filaments.  Similar characteristics of cold accretion
flows have been noted and discussed by \citet{Nagai03a},
\citet{Keres05}, and \citet{Dekel06}.

Entropy in the exterior of the halo differ little between adiabatic
and cooling runs outside of \rcool.  But as the gas falls within
\rcool, it cools and condenses, which gives a lower entropy, and the
$r$-$K$ histograms and entropy slices display this clearly.  Another
significant difference in the cooling simulations is the contraction
of the virial shock by a factor of 1/3 when compared to adiabatic
runs.  This is caused by the contraction of the cooling gas.  Here the
cold filaments penetrate to even smaller radii.  This is also evident
in the radial density profiles of Figure \ref{fig:dens}.

%
%
\begin{figure*}
\plottwo{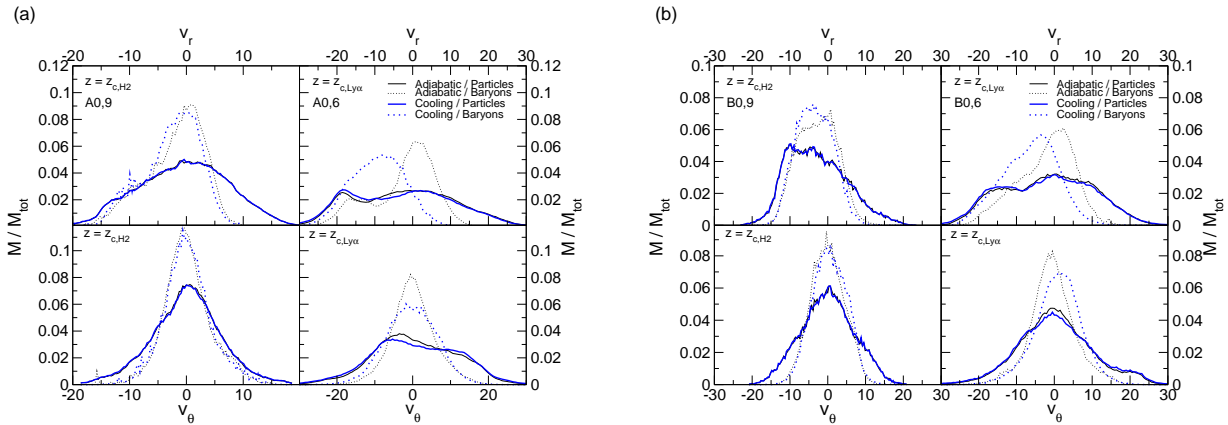}{f4b_color.eps}
\caption{\label{fig:veldistA} (a) simulation A. (b) simulation B.
  Radial (\textit{top}) and tangential (\textit{bottom}) velocity
  distributions of the most massive halo at z = \zhh~(\textit{left})
  and z = \zlya~(\textit{right}).  The \textit{heavy, blue} lines are
  the distributions of the adiabatic models, and the \textit{light,
    black} lines are from the radiative cooling models.
  \textit{Solid} and \textit{dotted} lines correspond to the velocity
  distributions of DM and baryons, respectively.  These distributions
  can be decomposed into single or multiple Gaussians, depending on
  substructure.  This demonstrates that violent relaxation occurs for
  the baryons as well as the DM. The narrower distributions of the
  baryons is due to the dissipation in shocks.}
\end{figure*}

\subsection{Virial Heating and Turbulence}
\label{sec:virial_heat_turb}

In order for a system to remain in virial equilibrium as it grows in
mass, additional gravitational energy is balanced through two possible
mechanisms: heating of the gas ($\dot{\mathcal{E}} > 0$) and
increasing the kinetic energy of the gas ($\dot{\mathcal{T}} > 0$).
We differentiate between two main cases of virialization by comparing
the cooling time, \tcool~= kT/n$\Lambda$, of the system to the heating
time, \theat~= \tvir/$\dot{T}_{\rm{vir}} \sim \frac{3}{2}
M_{\rm{vir}} / \dot{M}_{\rm{vir}}$ in the case of rapid mass
accretion.  \citet{Birnboim03}, \citet{Dekel06}, and \citet{Wang07}
find that radiative cooling rates are greater than heating rates from
virialization for halos with masses below $10^{12} \Ms$.

\medskip

1. \textit{Thermalization} (\tcool~$>$ \theat)--- When no efficient
radiative cooling mechanisms (e.g. \hh, \lya, \ion{He}{1}) exist, the
system virializes by injecting energy into heat $\mathcal{E}$ and
partly into kinetic energy $\mathcal{T}$.  In the process, the halo
becomes pressure supported and virialized.  Traditional galaxy
formation scenarios only consider this thermalization while neglecting
the kinetic energy term of equation (\ref{vt}).  However it is
important to regard kinetic energy, even in adiabatic models, as the
gas violently relaxes.  Turbulence velocities are similar to the
velocity dispersion of the system and contributes notably to the
overall energy budget as seen in adiabatic cases in Figure
\ref{fig:energies}.

\medskip

2. \textit{Turbulence generation} (\tcool~$<$ \theat)--- When a
cooling mechanism becomes efficient, the system now dispenses its
thermal energy and loses pressure support within \rcool.  The gas
will cool to a minimum equilibrium temperature.  As the
cooling halo collapses and radial velocities increase, the gas still
lacks enough kinetic and thermal energy to match the gravitational
energy and surface term in equation (\ref{vt}).  The gas becomes more
turbulent in order to virialize.  We see this in the second row of
Figure \ref{fig:energies}, where turbulent energies are significantly
increased well inside the halo in the cooling models as compared to
the adiabatic calculations.

\medskip

Through virial analyses, we have shown that turbulent energies are
comparable, if not dominant, to thermal energies in galaxy formation.
In the next Section, we further investigate the significance and
nature of the turbulence through velocity distributions and
decompositions in order to study any small-scale anisotropies in the
internal flows.

%
%
\begin{figure*}
\begin{center}
\plotone{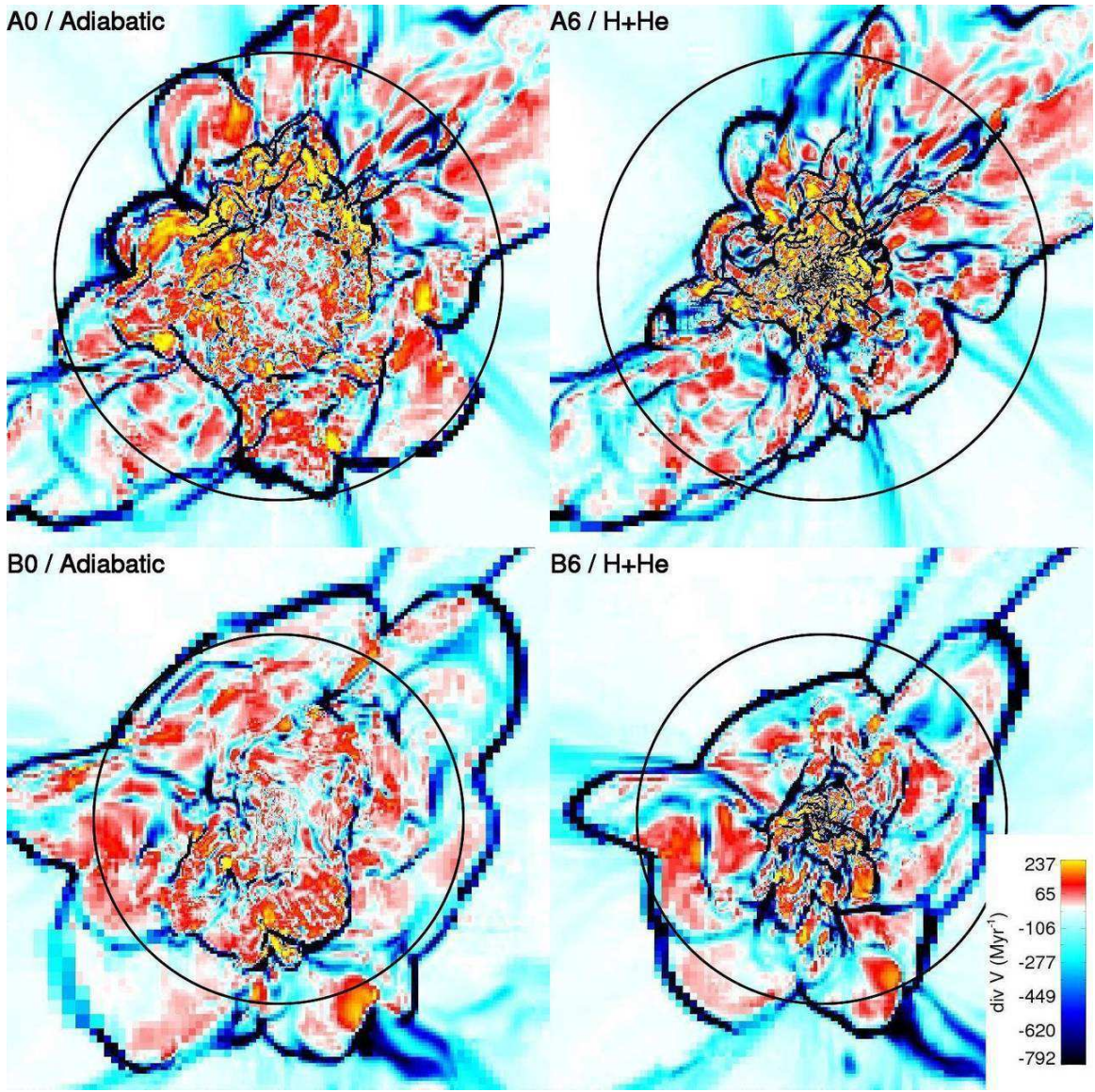}
\caption{\label{fig:divV} Two-dimensional slices of velocity
  divergence ($\nabla \cdot \vv$) at z = \zlya~for simulations A0, A6,
  B0, and B6.  The fields of view are 1.49 and 1.69 kpc for
  simulations A and B, respectively.  The circles mark the radius
  $r_{200}$ = 615 (576) pc for simulation A (B).  Shocks are clearly
  denoted by large, negative convergence values.  In the adiabatic
  cases, these shocks mainly exist at large radii where the gas from
  the voids and filaments virializes.  When we consider radiative
  cooling, supersonic turbulence increases the frequency of shock
  fronts in the interior of the halo.}
\end{center}
\end{figure*}

%
%

\begin{figure*}
\begin{center}
\plotone{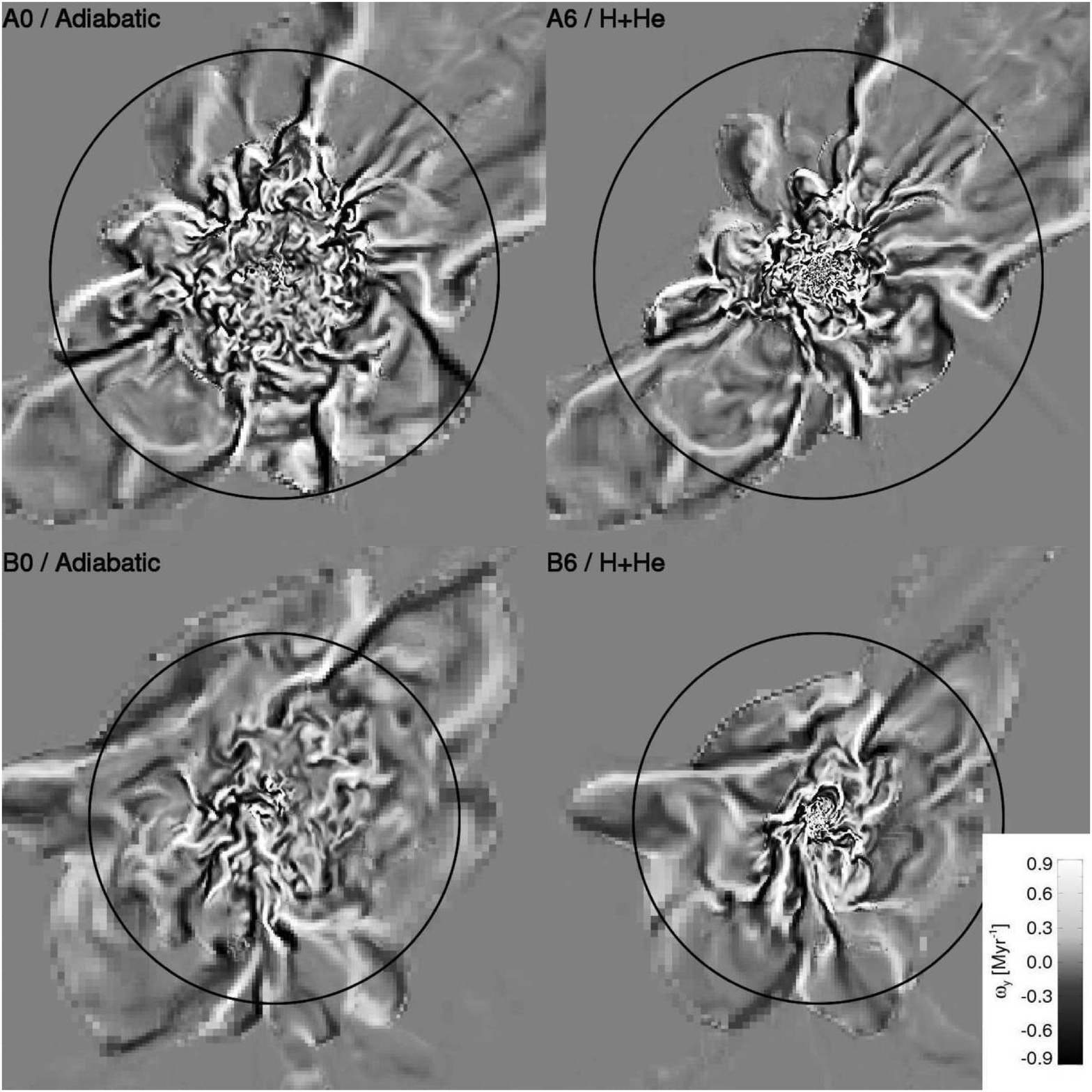}
\caption{\label{fig:curlV} Two-dimensional slices of the component
  perpendicular to the slice of vorticity ($\nabla \times \vv$) at z =
  \zlya~for simulations A0, A6, B0, and B6.  The fields of view and
  circle radii are the same as in Figure \ref{fig:divV}.  This
  quantity emphasizes the large- and small-scale turbulent eddies in
  the halo.}
\end{center}
\end{figure*}

\section{VELOCITY DISTRIBUTIONS}

In CDM cosmogony, collisionless dark matter dominates the
gravitational potential and oscillates as it stabilizes.  \citet{LB67}
showed how a collisionless system undergoes violent relaxation if
embedded within a rapidly time-varying potential.  Individual mass
elements do not conserve energy during violent relaxation, only the
entire system conserves energy.  This behavior randomizes the energies
of the mass elements, and statistical mechanics makes the resulting
energy (velocity) distribution to tend to Maxwellian.  Furthermore,
the system ``forgets'' its original configuration during virialization
or the incorporation of a lesser halo.  Later studies have inferred
two baryonic scenarios of virialization. First, violent relaxation and
the accompanying phase mixing also applies to the gaseous component
\citep{vdBosch02, Sharma05}.  In the other case, the gaseous component
dissipates all turbulent motions and finally rigidly rotates as a
solid body with a velocity appropriate to the overall spin parameter
\citep[e.g.][]{Loeb94, Mo98, Bromm03}.

Figure \ref{fig:veldistA} shows the velocity distributions of the dark
matter and baryonic components of the halo at \zhh~in the left column
and \zlya~in the right column.  It also overplots the simulations with
adiabatic and radiative cooling.  We plot the radial and tangential
velocity distribution on the top and bottom rows, respectively.  The
velocities are taken with respect to the bulk velocity of the halo.
We also transform the velocity components to align the z-axis and
total angular momentum vector of the DM halo.

The radial velocity distributions at z = \zhh~are approximately
Maxwellian in both dark matter and gas with a skew toward infall.  The
infall distributions are shifted by $\sim$1\kms~in the cooling case
when compared to adiabatic.  However at z = \zlya, the effects of
\lya~cooling become more prevalent in the halo when compared to
\hh~cooling, shifting the radial velocity distribution by
$\sim$5\kms~that is caused by faster infall.  These distributions have
two components that represent virialized gas and infalling gas in
filaments.  We further discuss this in the next Section.

The tangential velocity distributions are nearly Maxwellian in all
cases except for the dark matter in simulation A at z = \zlya~(right
panels in Figure \ref{fig:veldistA}a).  This deviation from Maxwellian
arises from two major mergers that occur between 25 and 85 Myr (z =
17--21) before the final collapse.  The residual substructure from the
major merger causes three distinct populations with Gaussian
distributions centered at --0.2, +13.6, and --6.7\kms~with $\sigma$
= 11.6, 4.2 and 3.6\kms, respectively.  These distributions clearly
do not resemble a solid body rotator, whose velocity distribution
would contain all positive velocities.  In other words the turbulent
velocities exceed the typical rotational speeds.

Distributions in dark matter are broader than the gas in both
simulations and collapse redshifts as expected because for the gas we
only give the bulk velocities and do not add the microscopic
dispersion \citep[cf.][]{vdBosch02, Sharma05}.

\subsection{Halo and Filament Contrasts}

The dark matter velocity distributions are typical of a virialized
system with the majority of the matter having a Maxwellian
distribution with a dispersion corresponding to the main halo
\citep{Boylan04, Dieman04, Kazantzidis04}.  Substructure appears as
smaller, superposed Gaussians, which are stripped of its outer
material as it orbits the parent halo.  Dynamical friction acts on the
substructure and decreases its pericenter over successive orbits, and
the subhalo is gradually assimilated in the halo.

The filaments penetrate deep into the halo and provide mostly radial
infall inside \rr.  They do not experience a virial shock at \rr, and
this contrast is apparent in the radial velocity distributions.  When
we restrict our analysis scope to the filaments (i.e. $r > 150$ pc and
$K < 10^{3.3}$ K cm$^{-2}$), the radial velocity distribution (Figure
\ref{fig:veldistA}) is skewed toward infall, centered at --15 km/s,
which is approximately the circular velocity of the halo.  The rest of
the gas outside of this region in $r-K$ space has already been
virialized, shock heated, and roughly exhibits a Maxwellian
distribution, centered at zero, with its associated substructures.
Hence the mass in filaments dominate the radial velocity distributions
at negative values in Figure \ref{fig:veldistA}.

\subsection{Turbulence}

Radial inflows can create turbulence in the halo.  Filaments provide
an influx material with distinct angular momentum.  This gas
virializes in the presence of an already turbulent medium that has a
relatively high specific angular momentum at $r > r_{200}/4$.  The
Rayleigh inviscid instability criterion requires
\begin{equation}
\label{eqn:rayleigh}
  \frac{dj^2}{dr} > 0 \quad \textrm{for rotational stability,}
\end{equation}
where $j$ is the specific angular momentum.  If this is not satisfied,
the system will become unstable to turbulence.  The onset of
turbulence can be delayed if viscosity were large enough so that
Reynolds numbers are below the order of 10$^2$ or 10$^3$.  However
there are many modes of instability if equation (\ref{eqn:rayleigh})
is not met, and even a gas with low Reynolds number will eventually
become fully turbulent \citep{Shu92}.

Velocity dispersions can characterize the general magnitude of a
turbulent medium, but its local nature is better detailed by applying
the Cauchy-Stokes decomposition,
\begin{equation}
  \uv^\prime = \uv + \frac{1}{2} \: {\bf \omega} \times \hv +
  \frac{1}{3} \: (\nabla \cdot \uv)\hv + 
  \frac{1}{2} \: \mathcal{D} \cdot \hv 
\end{equation}
that decomposes the velocity field into bulk motion $\uv$, vorticity
${\bf \omega} = \nabla \times \uv$, expansion and contraction $\nabla
\cdot \uv$, and a distortion $\mathcal{D}$ without change in volume.
Here the vector $\hv$ describes the separation between gas parcels at
position $\xv$ and $\xv^\prime$, and $\uv^\prime$ is the velocity at
$\xv^\prime$.

We relate $\nabla \cdot \uv$, which is plotted in Figure
\ref{fig:divV}, to a convergence timescale through the continuity
equation,
\begin{equation}
\label{eqn:continuity}
\dot{\rho} + \nabla(\rho\uv) = 0,
\end{equation}
that can be rewritten in terms of the total derivative $D/Dt$ as
\begin{equation}
\label{eqn:lct}
\frac{1}{\rho} \frac{D\rho}{Dt} = - \nabla \cdot \uv .
\end{equation}
$D\rho/Dt$ describes density changes along the fluid flow lines, and
the $1/\rho$ factor converts this change into an inverse timescale on
which local densities e-fold.  We denote $-(\nabla \cdot \uv)^{-1}$ as
the ``Lagrangian convergence timescale'' (LCT).  Converging flows
($\nabla \cdot \uv < 0$) are ubiquitous within the halo.  On large
scales, the smallest LCTs on the order of 20 kyr exist at the virial
shock, adjacent to both the filaments and voids.  In the cooling
models, these timescales are also small in turbulent shocks well
within \rr.  Analogous to the dynamical time, typical shocked LCTs
decrease toward the center as density increases.

The local magnitude and nature of turbulence is further illustrated by
the vorticity ${\bf \omega}$, whose component perpendicular to the
slice, $\omega_y$, is shown in Figure \ref{fig:curlV}.  The local
rotation period, $4\pi / \vert \omega \vert$ is also helpful to
quantify and visualize the nature of the flow.  The vorticity equation
reads
\begin{equation}
\frac{\partial\omega}{\partial t} + \nabla \times (\omega \times \uv) = 
\frac{1}{\rho^2} \nabla\rho \times \nabla p ,
\end{equation}
where the source term is non-zero when the density and pressure
gradients are not aligned, i.e. baroclinic vorticity generation.  This
occurs at and near shocks throughout the halo, regardless of radiative
cooling.  In the adiabatic models, vorticity exists even at modest but
sufficient resolution in the large pressure-supported cores (see
\mturb~for $r < 100$ pc in Figure \ref{fig:energies}) and generates a
turbulent medium with \mturb~= 0.3.  In the cooling models, this
large-scale vorticity is still present but increases in the collapsing
core.  As shocks become abundant in the center, we do not see any
dampening of kinetic energy.  Perhaps this mechanism maintains
turbulent motions during virialization, even in the presence of
dissipative shocks.  In Figure \ref{fig:curlV}, adjacent, antiparallel
fluid flows, i.e. a sign change in $\omega_y$, are ubiquitous, which
visually demonstrates that turbulence exists throughout the halo.  The
length scale of these eddies decrease with increasing density as with
the LCTs.

Hence we believe significant turbulence generated during virialization
should be present in all cosmological halos.  The cooling efficiency
of the gas, the total halo mass, and partly the merger history
determines the magnitude of turbulence.  We discuss some implications
of virial turbulence in the following section.

\section{DISCUSSION}

We have investigated the virialization of early pre-galactic
cosmological halos in this paper with a suite of AMR simulations with
varying chemistry and cooling models and collapse epochs.  When
analyzing the local virial equilibrium of the halo, we do not assume
that it is in equilibrium but explicitly calculate all of the relevant
terms in the virial theorem.  In both adiabatic and radiative cooling
cases, we find that the kinetic (turbulent) energy is comparable, if
not dominant, with the thermal energy.  Turbulence appreciates as
radiative cooling becomes efficient because thermal energy alone
cannot bring the system into virial equilibrium.  In this case, the
gas attempts to virialize by increasing and maintaining its kinetic
energy.

Besides violent relaxation, at least two other hydrodynamic processes
will augment virial turbulence.  The first occurs when radial inflow
interacts with the virialized gas.  Due to the Rayleigh criterion, the
high angular momentum gas creates an instability when it is deposited
by filaments at small radii.  The second happens when minor and major
mergers create Kelvin-Helmholtz instabilities and drives additional
turbulence \citep[e.g.][]{Ricker01, Takizawa05}.  Our results show
that this turbulence is acting to achieve close to virial equilibrium
at all stages during assembly and collapse.

Virial turbulence may be most important in halos which can cool
rapidly when compared to virial heating from mass accretion.
Interestingly all halo masses below $\sim$10$^{12} \Ms$ that can cool by
\lya~emission satisfy this condition \citep{Birnboim03, Dekel06,
  Wang07}.

Turbulence appears to mix angular momentum efficiently so that it
redistributes to a radially increasing function, and thus only the
lowest specific angular momentum material sinks to the center.  This
segregation allows a collapse to proceed as it were self-similar and
basically non-rotating.  Similar results have been reported in
cosmological simulations of collapses of the first stars
\citep{Abel00, Abel02, Yoshida03, Yoshida06b, OShea07}.  Here and in
protogalactic collapses, the turbulent velocities become supersonic.
One would expect even higher Mach numbers in larger potential wells
that still have \tcool~$<$~\tdyn.

The inclusion of the surface term allows us to study the virial
equilibrium in the halo's interior where the gravitational potential
is not influential.  Here our simulations show thermal and kinetic
energies balancing the surface term and potential energy to achieve
virial equilibrium.  Before cooling is efficient, gas virially heats
and its temperature can exceed the traditional virial temperature
within \rr/2~as seen in our adiabatic simulations.  The consequences
of this additional heating is substantial because halos can collapse
and form stars before the virial temperature reaches the critical
temperature, such as $\sim$10,000K for \lya~cooling.

The timescale at which the center collapse occurs is crucial to the
type of object that forms there.  If the collapse occurs faster than a
Kelvin-Helmholtz time for a massive star ($\sim$ 300 kyr), a black
hole might form from the lowest angular momentum gas.  Conversely if
the collapse is delayed by turbulent pressure, star formation could
occur in the density enhancements created by turbulent shocks.  The
ensuing radiative feedback may create outflows and thus slow further
infall and possibly prevent the formation of a central black hole.
The nature of the first galaxies poses an important question in the
high-redshift structure formation, and to address this problem we must
consider their progenitors -- the first stars.

\subsection{Pop III Feedback}

Numerical simulations have shown that the first, metal-free (Pop III)
stars form in isolation in its host halo.  They are believed to have
stellar masses $\sim$100 \Ms~\citep{Abel02, Omukai03, Tan04,
  Yoshida06b} and produce $\sim$$10^{50}$ photons s$^{-1}$ that can
ionize hydrogen and dissociate \hh~\citep{Schaerer02}.
One-dimensional radiative hydrodynamical calculations \citep{Whalen04,
  Kitayama04} and recently three-dimensional radiative hydrodynamical
AMR \citep{Abel07} and SPH \citep{Yoshida06a} simulations found that
pressure forces from the radiatively heated gas drive a
$\sim$30\kms~shock outwards and expels the majority of the gas in the
host halo.  Additionally the star ionizes the surrounding few kpc of
the intergalactic medium (IGM).

Pop III stellar feedback invalidates some of our assumptions in the
calculations presented here, but the general aspect of kinetic energy
being dominant should hold in the presence of these feedback
processes.  In a later paper, we will expand our simulations to
include radiative feedback from primordial stars \citep[cf.][]{
  Yoshida06a, Abel07} and the metal enrichment from pair instability
supernovae \citep{Barkat67, Bond84, Heger02} of the IGM and subsequent
star formation.

\section{SUMMARY}

We have investigated the process of virialization in pre-galactic gas
clouds in two cosmology AMR realizations.  Our virial analyses
included the kinetic (turbulent) energies and surface pressures of the
baryons in the system.  The significance of each energy component of
the gas varies with the effectiveness of the radiative cooling, which
we quantify by performing each realization with adiabatic, hydrogen
and helium, and \hh~cooling models.  We highlight the following main
results of this study as:

\medskip

1. Inside \rcool, gas cannot virialize alone through heating but must
gain kinetic energy.  It is up to a factor of five greater than thermal
energy throughout the protogalactic halos.  This manifests itself in a
faster bulk inflow and supersonic turbulent motions.

2. In the radiative cooling models, supersonic turbulence
($\mathcal{M}$ = 1--3) leads to additional cooling within turbulent
shocks.  We expect turbulence in larger galaxies, up to 10$^{12} \Ms$,
to be even more supersonic.

3. Baryonic velocity distributions are Maxwellian that shows violent
relaxation occurs for gas as well as dark matter.  Turbulent
velocities exceed typical rotational speeds, and these halos are
only poorly modeled as solid body rotators.

4. Virial shocks between the void-halo interface occur between
\rr/2~and \rr.  Dense, cold flows in filaments do not
shock-heat until well within \rr~and as small as \rr/4.

5. Turbulence generated during virialization mixes angular momentum so
that it redistributes to a radially increasing function (the Rayleigh
criterion).

\medskip

After the halo virializes, its central part will undergo turbulent
collapse, such as in primordial star formation and galactic molecular
clouds.  These collapses should be ubiquitous in early structure
formation as turbulence can be generated through virialization,
merging, and angular momentum segregation.  We conclude that
\textit{turbulence plays a key role in virialization and galaxy
  formation.}

\acknowledgments{This work was supported by NSF CAREER award
  AST-0239709 from the National Science Foundation.  We appreciate
  Marcelo Alvarez and an anonymous referee providing very constructive
  feedback on this paper.  We thank Michael Norman, Ralph Pudritz,
  Darren Reed, Tom Theuns, and Peng Wang for helpful discussions.  We
  also thank Paul Kunz for his invaluable help with our analysis
  software.  We are grateful for the continuous support from the
  computational team at SLAC.  We performed these calculations on 16
  processors of a SGI Altix 3700 Bx2 at KIPAC at Stanford University.}

{}


\begin{thebibliography}{}


\bibitem[\protect\citeauthoryear{Abel et al.}{1997}]{Abel97} Abel, T.,
  Anninos, P., Zhang, Y., \& Norman, M.~L.\ 1997, New Astronomy, 2,
  181

\bibitem[\protect\citeauthoryear{Abel et al.}{2000}]{Abel00} Abel, T.,
  Bryan, G.~L., \& Norman, M.~L. 2000, \apj, 540, 39

\bibitem[\protect\citeauthoryear{Abel et al.}{2002}]{Abel02}
  ---. 2002, Science, 295, 93

\bibitem[\protect\citeauthoryear{Abel et al.}{2007}]{Abel07} Abel, T.,
  Wise, J.~H., \& Bryan, G.~L. 2007, \apjl, 659, L87

\bibitem[\protect\citeauthoryear{Alvarez et al.}{2006}]{Alvarez06}
  Alvarez, M.~A., Bromm, V., \& Shapiro, P.~R.\ 2006, \apj, 639, 621

\bibitem[\protect\citeauthoryear{Anninos et al.}{1997}]{Anninos97}
  Anninos, P., Zhang, Y., Abel, T., \& Norman, M.~L.\ 1997, New
  Astronomy, 2, 209

\bibitem[\protect\citeauthoryear{Ballesteros-Paredes}{2006}]{Ballesteros06}
  Ballesteros-Paredes, J.\ 2006, \mnras, 372, 443

\bibitem[\protect\citeauthoryear{Barkat et al.}{1967}]{Barkat67}
  Barkat, Z., Rakavy, G., \& Sack, N.\ 1967, Physical Review Letters,
  18, 379


\bibitem[\protect\citeauthoryear{Begelman et al.}{2006}]{Begelman06}
  Begelman, M.~C., Volonteri, M., \& Rees, M.~J.\ 2006, \mnras, 370,
  289

\bibitem[\protect\citeauthoryear{Bertschinger}{1995}]{Bertschinger95}
  Bertschinger, E.\ 1995, ArXiv Astrophysics e-prints,
  arXiv:astro-ph/9506070

\bibitem[\protect\citeauthoryear{Bertschinger}{2001}]{Bertschinger01}
  Bertschinger, E.\ 2001, \apjs, 137, 1

\bibitem[\protect\citeauthoryear{Birnboim \& Dekel}{2003}]{Birnboim03}
  Birnboim, Y., \& Dekel, A.\ 2003, \mnras, 345, 349

\bibitem[\protect\citeauthoryear{Blumenthal et
    al.}{1984}]{Blumenthal84} Blumenthal, G.~R., Faber, S.~M.,
  Primack, J.~R., \& Rees, M.~J.\ 1984, \nat, 311, 517

\bibitem[\protect\citeauthoryear{Bond et al.}{1984}]{Bond84} Bond,
  J.~R., Arnett, W.~D., \& Carr, B.~J.\ 1984, \apj, 280, 825

\bibitem[\protect\citeauthoryear{Bond et al.}{1991}]{Bond91} Bond,
  J.~R., Cole, S., Efstathiou, G., \& Kaiser, N.\ 1991, \apj, 379, 440

\bibitem[\protect\citeauthoryear{Bower}{1991}]{Bower91} Bower, R.~G.\
  1991, \mnras, 248, 332

\bibitem[\protect\citeauthoryear{Boylan-Kolchin \& Ma}{2004}]
  {Boylan04} Boylan-Kolchin, M., \& Ma, C.-P.\ 2004, \mnras, 349, 1117

\bibitem[\protect\citeauthoryear{Bromm \& Loeb}{2003}]{Bromm03} Bromm,
  V. \& Loeb, A, 2003, \apj, 596, 34


\bibitem[\protect\citeauthoryear{Bryan et al.}{1995}]{Bryan95} Bryan,
  G.~L., Norman, M.~L., Stone, J.~M., Cen, R., Ostriker, J.~P.\ 1995,
  Computer Physics Communication 89, 149

\bibitem[\protect\citeauthoryear{Bryan \& Norman}{1997}]{Bryan97}
  Bryan, G.~L. \& Norman, M.~L. 1997, in Computational Astrophysics,
  eds. D.~A. Clarke and M. Fall, ASP Conference \#123

\bibitem[Bryan \& Norman(1998)]{Bryan98} Bryan, G.~L., \& 
  Norman, M.~L.\ 1998, \apj, 495, 80 
 
\bibitem[\protect\citeauthoryear{Bryan \& Norman}{1999}]{Bryan99}
  Bryan, G.~L. \& Norman, M.~L. 1999, in Workshop on Structured
  Adaptive Mesh Refinement Grid Methods, IMA Volumes in Mathematics
  No. 117, ed. N. Chrisochoides, p. 165

\bibitem[\protect\citeauthoryear{Burles et al.}{2001}]{Burles01}
  Burles, S., Nollett, K.~M., \& Turner, M.~S.\ 2001, \apjl, 552, L1

\bibitem[\protect\citeauthoryear{Cen}{2005}]{Cen05} Cen, R.\ 2005,
  \apj, 620, 191

\bibitem[\protect\citeauthoryear{Chandrasekhar}{1961}]{Chandra61}
  Chandrasekhar, S.\ 1961, \textit{Hydrodynamic and hydromagnetic
    stability}, (Oxford: Clarendon)

\bibitem[\protect\citeauthoryear{Chandrasekhar \& Fermi}{1953}]
  {Chandra53} Chandrasekhar, S., \& Fermi, E.\ 1953, \apj, 118, 116

\bibitem[\protect\citeauthoryear{Cole et al.}{1994}]{Cole94} Cole, S.,
  Aragon-Salamanca, A., Frenk, C.~S., Navarro, J.~F., \& Zepf, S.~E.\
  1994, \mnras, 271, 781

\bibitem[\protect\citeauthoryear{Cole et al.}{2000}]{Cole00} Cole, S.,
  Lacey, C.~G., Baugh, C.~M., \& Frenk, C.~S.\ 2000, \mnras, 319, 168

\bibitem[\protect\citeauthoryear{Couchman}{1991}]{Couchman91}
Couchman, H.~M.~P.\ 1991, \apjl, 368, L23

\bibitem[\protect\citeauthoryear{Crampin \& Hoyle}{1964}]{Crampin64}
  Crampin, D.~J., \& Hoyle, F.\ 1964, \apj, 140, 99

\bibitem[\protect\citeauthoryear{De Lucia et al.}{2004}]{DeLucia04}
  De Lucia, G., Kauffmann, G., Springel, V., White, S.~D.~M., Lanzoni,
  B., Stoehr, F., Tormen, G., \& Yoshida, N.\ 2004, \mnras, 348, 333

\bibitem[\protect\citeauthoryear{Dekel \& Birnboim}{2006}]{Dekel06}
  Dekel, A., \& Birnboim, Y.\ 2006, \mnras, 368, 2

\bibitem[\protect\citeauthoryear{Dieman et al.}{2004}]{Dieman04}
  Diemand, J., Moore, B., \& Stadel, J.\ 2004, \mnras, 352, 535

\bibitem[\protect\citeauthoryear{Dolag et al.}{2004}]{Dolag04} Dolag,
  K., Jubelgas, M., Springel, V., Borgani, S., \& Rasia, E.\ 2004,
  \apjl, 606, L97

\bibitem[\protect\citeauthoryear{Dolag et al.}{2005}]{Dolag05} Dolag,
  K., Vazza, F., Brunetti, G., \& Tormen, G.\ 2005, \mnras, 364, 753

\bibitem[\protect\citeauthoryear{Draine \& Bertoldi}{1996}]{Draine96}
  Draine, B.~T., \& Bertoldi, F.\ 1996, \apj, 468, 269

\bibitem[\protect\citeauthoryear{Eisenstein \&
    Hut}{1998}]{Eisenstein98} Eisenstein, D.~J.~\& Hut, P.\ 1998,
  \apj, 498, 137

\bibitem[\protect\citeauthoryear{Fall \& Efstathiou}{1980}]{Fall80}
  Fall, S.~M.~\& Efstathiou, G.\ 1980, \mnras, 193, 189

\bibitem[\protect\citeauthoryear{Flower et al.}{2000}]{Flower00}
  Flower, D.~R., Le Bourlot, J., Pineau des For{\^e}ts, G., \& Roueff,
  E.\ 2000, \mnras, 314, 753

\bibitem[\protect\citeauthoryear{Galli \& Palla}{1998}]{Galli98}
  Galli, D., \& Palla, F.\ 1998, \aap, 335, 403

\bibitem[\protect\citeauthoryear{Gao et al.}{2005}]{Gao05} Gao, L.,
  White, S.~D.~M., Jenkins, A., Frenk, C.~S., \& Springel, V.\ 2005,
  \mnras, 363, 379

\bibitem[\protect\citeauthoryear{Goldman}{2000}]{Goldman00} Goldman,
  I.\ 2000, \apj, 541, 701


\bibitem[\protect\citeauthoryear{Heger \& Woosley}{2002}]{Heger02}
  Heger, A.~\& Woosley, S.~E.\ 2002, \apj, 567, 532

\bibitem[\protect\citeauthoryear{Hu \& Dodelson}{2002}]{Hu02} Hu,
  W.~\& Dodelson, S.\ 2002, \araa, 40, 171

\bibitem[\protect\citeauthoryear{Iliev \& Shapiro}{2001}]{Iliev01}
  Iliev, I.~T., \& Shapiro, P.~R.\ 2001, \mnras, 325, 468


\bibitem[\protect\citeauthoryear{Kazantzidis et al.}{2004}]
  {Kazantzidis04} Kazantzidis, S., Magorrian, J., \& Moore, B.\ 2004,
  \apj, 601, 37

\bibitem[\protect\citeauthoryear{Kennicutt}{1998}]{Kennicutt98}
  Kennicutt, R.~C., Jr.\ 1998, \araa, 36, 189

\bibitem[\protect\citeauthoryear{Kere{\v s} et al.}{2005}]{Keres05}
  Kere{\v s}, D., Katz, N., Weinberg, D.~H., \& Dav{\'e}, R.\ 2005,
  \mnras, 363, 2

\bibitem[\protect\citeauthoryear{Kim \& Narayan}{2003}]{Kim03} Kim,
  W.-T., \& Narayan, R.\ 2003, \apj, 596, 889

\bibitem[\protect\citeauthoryear{Kitayama et al.}{2004}]{Kitayama04}
  Kitayama, T., Yoshida, N., Susa, H., \& Umemura, M.\ 2004, \apj,
  613, 631

\bibitem[\protect\citeauthoryear{Kuhlen \& Madau}{2005}]{Kuhlen05}
  Kuhlen, M., \& Madau, P.\ 2005, \mnras, 363, 1069

\bibitem[\protect\citeauthoryear{Lacey \& Cole}{1993}]{Lacey93} Lacey,
  C., \& Cole, S.\ 1993, \mnras, 262, 627

\bibitem[\protect\citeauthoryear{Lacey \& Cole}{1994}]{Lacey94} Lacey,
  C., \& Cole, S.\ 1994, \mnras, 271, 676

\bibitem[\protect\citeauthoryear{Lacey \& Silk}{1991}]{Lacey91} Lacey,
  C., \& Silk, J.\ 1991, \apj, 381, 14

\bibitem[\protect\citeauthoryear{Larson}{1979}]{Larson79} Larson,
  R.~B.\ 1979, \mnras, 186, 479

\bibitem[\protect\citeauthoryear{Larson}{1981}]{Larson81} Larson,
  R.~B.\ 1981, \mnras, 194, 809

\bibitem[\protect\citeauthoryear{Larson}{2003}]{Larson03} Larson,
  R.~B.\ 2003, Reports of Progress in Physics, 66, 1651

\bibitem[\protect\citeauthoryear{Loeb \& Rasio}{1994}]{Loeb94} Loeb,
  A., \& Rasio, F.~A.\ 1994, \apj, 432, 52

\bibitem[\protect\citeauthoryear{Lynden-Bell}{1967}]{LB67}
  Lynden-Bell, D.\ 1967, \mnras, 136, 101


\bibitem[\protect\citeauthoryear{Mo et al.}{1998}]{Mo98} Mo, H.~J.,
  Mao, S., \& White, S.~D.~M.\ 1998, \mnras, 295, 319

\bibitem[\protect\citeauthoryear{Myers}{1999}]{Myers99} Myers, P.~C.\
  1999, in The Physics and Chemistry of the Interstellar Medium, eds.
  V.~Ossenkopf, J.~Stutzki, and G.~Winnewisser, 3rd Cologne-Zermatt
  Symposium, p. 227

\bibitem[\protect\citeauthoryear{Nagai \& Kravtsov}{2003a}]{Nagai03a}
  Nagai, D., \& Kravtsov, A.~V.\ 2003a, \apj, 587, 514

\bibitem[\protect\citeauthoryear{Nagai et al.}{2003b}]{Nagai03b}
  Nagai, D., Kravtsov, A.~V., \& Kosowsky, A.\ 2003, \apj, 587, 524

\bibitem[\protect\citeauthoryear{Norman \& Bryan}{1999}]{Norman99}
  Norman, M.~L., \& Bryan, G.~L.\ 1999, LNP Vol.~530: The Radio Galaxy
  Messier 87, 530, 106


\bibitem[\protect\citeauthoryear{O'Shea et al.}{2004}]{OShea04}
  O'Shea, B.~W., Bryan, G., Bordner, J., Norman, M.~L., Abel, T.,
  Harkness, R., \& Kritsuk, A.\ 2004, Adaptive Mesh Refinement -
  Theory and Applications, eds. T.~Plewa, T.~Linde \& V.~G.~Weirs,
  arXiv:astro-ph/0403044

\bibitem[\protect\citeauthoryear{O'Shea \& Norman}{2007}]{OShea07}
  O'Shea, B.~W., \& Norman, M.~L.\ 2007, \apj, 654, 66



\bibitem[\protect\citeauthoryear{Omukai \& Palla}{2003}]{Omukai03}
  Omukai, K., \& Palla, F.\ 2003, \apj, 589, 677


\bibitem[\protect\citeauthoryear{Palla et al.}{1983}]{Palla83} Palla,
  F., Salpeter, E.~E., \& Stahler, S.~W.\ 1983, \apj, 271, 632
 

\bibitem[\protect\citeauthoryear{Ponman et al.}{1999}]{Ponman99}
  Ponman, T.~J., Cannon, D.~B., \& Navarro, J.~F.\ 1999, \nat, 397,
  135

\bibitem[\protect\citeauthoryear{Press \& Schechter}{1974}]{Press74}
  Press, W.~H.~\& Schechter, P.\ 1974, \apj, 187, 425

\bibitem[Rayleigh(1920)]{Rayleigh20} Lord Rayleigh.\ 1920,
  \textit{Scientific Papers}, 6, 447

\bibitem[\protect\citeauthoryear{Rees \& Ostriker}{1977}]{Rees77}
  Rees, M.~J., \& Ostriker, J.~P.\ 1977, \mnras, 179, 541

\bibitem[\protect\citeauthoryear{Ricker \& Sarazin}{2001}]{Ricker01}
  Ricker, P.~M., \& Sarazin, C.~L.\ 2001, \apj, 561, 621


\bibitem[\protect\citeauthoryear{Saslaw \& Zipoy}{1967}]{Saslaw67}
  Saslaw, W.~C., \& Zipoy, D.\ 1967, \nat, 216, 976

\bibitem[\protect\citeauthoryear{Schaerer}{2002}]{Schaerer02}
  Schaerer, D.\ 2002, \aap, 382, 28


\bibitem[\protect\citeauthoryear{Schuecker et al.}{2004}]{Schuecker04}
  Schuecker, P., Finoguenov, A., Miniati, F., B{\"o}hringer, H., \&
  Briel, U.~G.\ 2004, \aap, 426, 387

\bibitem[\protect\citeauthoryear{Shapiro et al.}{1994}]{Shapiro94}
  Shapiro, P.~R., Giroux, M.~L., \& Babul, A.\ 1994, \apj, 427, 25

\bibitem[\protect\citeauthoryear{Shapiro et al.}{1999}]{Shapiro99}
  Shapiro, P.~R., Iliev, I.~T., \& Raga, A.~C.\ 1999, \mnras, 307, 203

\bibitem[\protect\citeauthoryear{Sharma \& Steinmetz}{2005}]{Sharma05}
  Sharma, S., \& Steinmetz, M.\ 2005, \apj, 628, 21

\bibitem[\protect\citeauthoryear{Shaw et al.}{2006}]{Shaw06} Shaw,
  L.~D., Weller, J., Ostriker, J.~P., \& Bode, P.\ 2006, \apj, 646,
  815

\bibitem[\protect\citeauthoryear{Sheth \& Tormen}{2002}]{Sheth02}
  Sheth, R.~K., \& Tormen, G.\ 2002, \mnras, 329, 61

\bibitem[\protect\citeauthoryear{Shu}{1992}]{Shu92} Shu, F.~H.\ 1992,
  \textit{Physics of Astrophysics, Vol.~II}, (Sausalito: University
  Science Books)

\bibitem[\protect\citeauthoryear{Silk}{1977}]{Silk77} Silk, J.\ 1977,
  \apj, 211, 638

\bibitem[\protect\citeauthoryear{Spaans \& Silk}{2006}]{Spaans06}
  Spaans, M., \& Silk, J.\ 2006, \apj, 652, 902

\bibitem[\protect\citeauthoryear{Spergel et al.}{2003}]{Spergel03}
  Spergel, D.~N., et al.\ 2003, \apjs, 148, 175

\bibitem[\protect\citeauthoryear{Spergel et al.}{2006}]{Spergel06}
  Spergel, D.~N., et al.\ 2007, \apj, \textit{submitted},
  \texttt{(astro-ph/0603449)}

\bibitem[\protect\citeauthoryear{Springel et al.}{2005}]{Springel05}
  Springel, V., Di Matteo, T., \& Hernquist, L.\ 2005, \mnras, 361,
  776

\bibitem[\protect\citeauthoryear{Takizawa}{2005}]{Takizawa05}
  Takizawa, M.\ 2005, \apj, 629, 791

\bibitem[\protect\citeauthoryear{Tan \& McKee}{2004}]{Tan04} Tan,
  J.~C., \& McKee, C.~F.\ 2004, \apj, 603, 383

\bibitem[\protect\citeauthoryear{Truelove et al.}{1997}]{Truelove97}
  Truelove, J.~K., Klein, R.~I., McKee, C.~F., Holliman, J.~H.,
  Howell, L.~H., \& Greenough, J.~A.\ 1997, ApJL, 489, L179

\bibitem[\protect\citeauthoryear{Tumlinson}{2006}]{Tumlinson06}
  Tumlinson, J.\ 2006, \apj, 641, 1

\bibitem[\protect\citeauthoryear{van den Bosch}{2002}]{vdBosch02b} van
  den Bosch, F.~C.\ 2002, \mnras, 331, 98

\bibitem[\protect\citeauthoryear{van den Bosch et al.}{2002}]
  {vdBosch02} van den Bosch, F.~C., Abel, T., Croft, R.~A.~C.,
  Hernquist, L., \& White, S.~D.~M.\ 2002, \apj, 576, 21

\bibitem[\protect\citeauthoryear{Volonteri et al.}{2005}]{Volonteri05}
  Volonteri, M., \& Rees, M.~J.\ 2005, \apj, 633, 624

\bibitem[\protect\citeauthoryear{Wang \& Abel}{2007}]{Wang07} Wang,
  P., \& Abel, T.\ 2007, \apj, \textit{submitted},
  \texttt{(astro-ph/0701363)}

\bibitem[\protect\citeauthoryear{Whalen et al.}{2004}]{Whalen04}
  Whalen, D., Abel, T., \& Norman, M.~L.\ 2004, \apj, 610, 14

\bibitem[\protect\citeauthoryear{White \& Frenk}{1991}]{White91}
  White, S.~D.~M., \& Frenk, C.~S.\ 1991, \apj, 379, 52

\bibitem[White \& Rees(1978)]{White78} White, S.~D.~M., \& Rees,
  M.~J.\ 1978, \mnras, 183, 341


\bibitem[\protect\citeauthoryear{Woodward \&
    Colella}{1984}]{Woodward84} Woodward, P.~R. \& Colella, P.\ 1984,
  J. Comput. Phys. 54, 115

\bibitem[\protect\citeauthoryear{Yoshida et al.}{2003}]{Yoshida03}
  Yoshida, N., Abel, T., Hernquist, L., \& Sugiyama, N.\ 2003, \apj,
  592, 645

\bibitem[\protect\citeauthoryear{Yoshida et al.}{2006a}]{Yoshida06a}
  Yoshida, N., Oh, S.~P., Kitayama, T., \& Hernquist, L.\ 2006a,
  \mnras, \textit{submitted}, \texttt{(astro-ph/0610819)}

\bibitem[\protect\citeauthoryear{Yoshida et al.}{2006b}]{Yoshida06b}
  Yoshida, N., Omukai, K., Hernquist, L., \& Abel, T.\ 2006b, \apj,
  652, 6

\end{thebibliography}
\end{document}